\documentclass[]{bmcart}

\usepackage{amsthm,amsmath,amssymb}
\usepackage{bm}
\usepackage[utf8]{inputenc} 
\usepackage{graphicx}
\graphicspath{{./Figs/}}
\usepackage{cancel}
\usepackage{booktabs}
\usepackage{hyperref}
\usepackage{siunitx}

\newcommand*\diff{\mathop{}\!\mathrm{d}}
\usepackage{upgreek}
\usepackage{graphicx}
\usepackage{subfigure}
\usepackage{color}
\usepackage[normalem]{ulem} 
\newcommand{\Replace}[2]{\bgroup\noindent\textcolor{red}{\xout{#1} #2}\egroup\ignorespacesafterend}
\newcommand{\Delete} [1]{\bgroup\noindent\textcolor{red}{\xout{#1}}\egroup\ignorespacesafterend}
\newcommand{\Insert} [1]{\bgroup\noindent\textcolor{}{#1}\egroup\ignorespacesafterend}
\newcommand{\Comment}[1]{\definecolor{Mygray}{gray}{0.50}\bgroup\color{Mygray}\noindent#1\egroup\ignorespacesafterend}
\newcommand \Michael [1]{\bgroup\noindent[\textcolor{blue}{\textbf{Michael}: #1}]\egroup\ignorespacesafterend}
\newcommand \Jonas  [1]{\bgroup\noindent[\textcolor{blue}{\textbf{Jonas}: #1}]\egroup\ignorespacesafterend}

\DeclareMathAlphabet{\Ibb}{U}{msb}{m}{n}

\newcommand{\BC}{{\boldsymbol{\mathnormal C}}}

\newcommand{\BX}{{\boldsymbol{\mathnormal X}}}
\newcommand{\superscr}[1]{\ensuremath{{}^{\rm #1}}}

\newcommand{\Bxi}    {\ensuremath{\boldsymbol\xi}}
\newcommand{\Beta}   {\ensuremath{\boldsymbol\eta}}
\newcommand{\Bbeta}   {\ensuremath{\boldsymbol\beta}}

\newcommand{\Bsigma} {\ensuremath{\boldsymbol\sigma}}
\newcommand{\Bnabla} {\ensuremath{\boldsymbol\nabla}}

\newcommand{\Bbetapl}{\ensuremath{\boldsymbol\beta\superscr{pl}}}

\newcommand{\Bb}{{\boldsymbol{\mathnormal b}}}

\newcommand{\Be}{{\boldsymbol{\mathnormal e}}}
\newcommand{\Bf}{{\boldsymbol{\mathnormal f}}}
\newcommand{\Bg}{{\boldsymbol{\mathnormal g}}}

\newcommand{\Bn}{{\boldsymbol{\mathnormal n}}}

\newcommand{\Br}{{\boldsymbol{\mathnormal r}}}

\newcommand{\Bt}{{\boldsymbol{\mathnormal t}}}
\newcommand{\Bu}{{\boldsymbol{\mathnormal u}}}

\newcommand{\Bx}{{\boldsymbol{\mathnormal x}}}

\newcommand{\Balpha }{\ensuremath{\boldsymbol\alpha}}

\newcommand{\figref}[1]{Fig.~\ref{#1}}

\newcommand \MZ [1] {\bgroup\noindent[\textcolor{blue}{\textbf{MZ}: #1}]\egroup\ignorespacesafterend}
\newcommand \JR [1] {\bgroup\noindent[\textcolor{orange}{\textbf{JR}: #1}]\egroup\ignorespacesafterend}

\begin{document}

\begin{frontmatter}

\begin{fmbox}
\dochead{Research}


\title{Multiscale modeling of dislocations: Combining peridynamics with gradient elasticity}


\author[
   addressref={aff1},                   
]{\inits{JR}\fnm{Jonas} \snm{Ritter}}
\author[
   addressref={aff1},
   corref={aff1},                       
   email={michael.zaiser@fau.de}
]{\inits{MZ}\fnm{Michael} \snm{Zaiser}}


\address[id=aff1]{
  \orgname{Department of Materials Simulation, Friedrich-Alexander Universität Erlangen-Nürnberg}, 
  \street{Dr.-Mack-Str. 77},                     %
  \postcode{90762}                                
  \city{Fürth},                              
  \cny{Germany}                                    
}



\end{fmbox}


\begin{abstractbox}

\begin{abstract} 
Modeling dislocations is an inherently multiscale problem as one needs to simultaneously describe the high stress fields near the dislocation cores, which depend on atomistic length scales, and a surface boundary value problem which depends on boundary conditions on the sample scale. We present a novel approach which is based on a peridynamic dislocation model to deal with the surface boundary value problem. In this model, the  singularity of the stress field at the dislocation core is regularized owing to the non-local nature of peridynamics. The effective core radius is defined by the peridynamic horizon which, for reasons of computational cost, must be chosen much larger than the lattice constant. This implies that dislocation stresses in the near-core region are seriously underestimated. By exploiting relationships between peridynamics and Mindlin-type gradient elasticity, we then show that gradient elasticity can be used to construct short-range corrections to the peridynamic stress field that yield a correct description of dislocation stresses from the atomic to the sample scale. 
\end{abstract}




\end{abstractbox}

\end{frontmatter}

\section[Introduction]{Introduction}

Peridynamics as introduced by Silling \cite{silling2000reformulation} is a nonlocal continuum theory  based on the formulation of integrodifferential equations for the displacement field, without directly involving classical concepts such as stress and strain. Like other generalized continuum approaches such as nonlocal elasticity  \cite{eringen1983differential} or gradient elasticity 
\cite{mindlin1965second,mindlin1968first,kroner1967elasticity}, it can serve to regularize problems that, in classical elasticity theory, are associated with singular or discontinuous solutions. Peridynamics shares this regularization property with higher-order continuum theories such as gradient elasticity (see e.g. \cite{silling2008convergence}), and several authors have proposed matching schemes between peridynamics and higher-order continua \cite{dayal2017leading,chen2020higher}. 

Here we use bond-based peridynamics to explore how peridynamics can be used to model dislocations -- a paradigmatic multiscale problem as, in classical elasticity, the elastic fields of dislocations exhibit singularities both at infinity (where they are regularized by surface boundary conditions) and at the dislocation line (where they are regularized by atomic-scale non-local interactions). One thus needs to establish a consistent description of dislocation stress fields from the atomic to the sample scale. In section \ref{sec:2}, we give a brief introduction into bond-based peridynamics and derive its relations with bulk gradient elasticity. In particular, we show how to relate the micromodulus and length scale parameter ('horizon') of peridynamics to the elastic constants and higher-order elastic constants of Mindlin's first strain gradient elasticity. In section \ref{sec:disloc} the introduction of dislocations into a  peridynamic framework is discussed, using concepts of continuum dislocation theory which relate dislocations to spatial derivatives of the plastic distortion. We first consider singular dislocations which, in generalization of Volterra's construction, can be considered in terms of displacement discontinuities across slip planes; in this formulation, the plastic distortion localized on a slip plane leads to force-free changes in bond vectors that cross this plane. Similar ideas have been formulated previously by other authors, see in particular Zhao et. al. \cite{zhao2021nonlocal}; the formulation here differs from theirs in detail as the horizon of peridynamic interactions is defined in an intermediate (plastically deformed) configuration rather than the material reference configuration. This allows to correctly account for the physical implications of changes in atomic neighborhood relations.  The formulation is then generalized to dislocations of distributed Burgers vector, where accordingly the plastic distortion is spatially distributed around the dislocation slip plane and the force-free change of the bond vector must be evaluated as a line  integral over the plastic distortion along the bond. For both singular and distributed dislocations, it is shown that the dislocation related fields exhibit a peculiar scaling form which is common to the classical, singular solutions obtained from linear elasticity, to solutions computed in the framework of gradient elasticity, and to the results obtained numerically from peridynamics. 
Section \ref{sec:peridis} discusses numerical results obtained for singular as well as distributed edge dislocations in the bond-based peridynamic framework. For singular dislocations where the plastic distortion is localized on the slip plane, it is found that the peridynamic solution removes the stress and strain singularities at the dislocation line, but still exhibits non analytic behavior as the asymptotic solution, that is approached when the number of collocation points in the horizon is increased, exhibits a discontinuous jump across the dislocation line. It is then shown that this jump is removed when dislocations with distributed Burgers vector are considered, and numerical results obtained with appropriate Burgers vector distribution are shown to be in excellent agreement with theoretical results for dislocation stress fields in gradient elasticity. 
Finally, section \ref{sec:hybrid} shows how the results can be used to construct a multiscale description of dislocations akin to the FEM-DDD (Finite Element Method-Discrete Dislocation Dynamics) hybrid models that are often used in discrete dislocation simulations \cite{huang2015coupled,jamond2016consistent,lu2019grain,lu2022size}. Accordingly, here it is proposed to use  peridynamics for describing the elastic field far from the dislocation and the associated boundary value problems, while gradient elasticity results are used to provide corrections that provide an accurate description of the stresses close to the dislocation line.

\section{Structure of bond-based peridynamics and relations with gradient elasticity}
\label{sec:2} 

\subsection{Bond-based Peridynamics}\label{sec_bondbased}

In bond-based peridynamics model, as originally defined by Silling \cite{silling2000reformulation}, the deformation of a $D$ dimensional continuous body ${\cal B}$ of density $\rho(\Bx)$ is described by the displacement field $\Bu(\Bx)$ where $\Bx$ are material coordinates.  The force balance equation for the point $\Bx$ is given by
\begin{equation}
	\rho(\Bx) \ddot{\Bu}(\Bx) = \int_{{\cal H}_{\Bx}} \Bf(\Bx,\Bx') \diff \Bx' + \Bb(\Bx) ,
\end{equation}
where $\Bf(\Bx,\Bx')$ is a pair force density between $\Bx'$ and $\Bx$, $\Bb$ is a body force field, and interactions are restricted to an interaction domain ${\cal H}_{\Bx}$ which we take to be a $D$ dimensional sphere of radius $\delta_{\textrm{p}}$ around $\Bx$, the so-called horizon: $(|\Bx - \Bx^*| \le \delta_{\textrm{p}}) \; \forall \; \Bx^* \in {\cal H}_{\Bx}$. The pair force density is specified constitutively. To this end, the bond stretch is defined as
\begin{equation}
	s = \frac{\left\vert \Beta + \Bxi \right\vert - \left\vert \Bxi \right\vert}{\left\vert\Bxi\right\vert}
	\label{eq_bond_stretch}
\end{equation}
where $\Bxi = \Bx^{\prime} - \Bx$ is the bond vector and $\Beta = \Bu^{\prime} - \Bu$ is the relative bond displacement. The pair force density is then taken to be linearly proportional to the bond stretch $s$, and pointing in the direction of the vector $\Be_{\Bu}\left(\Bxi,\Beta \right)$ connecting both points in the current configuration:
\begin{equation}
	\Bf \left(\Bx,\Bx^{\prime} \right)= c \left(\Bxi \right) s \Be_{\Bu}\left(\Bxi,\Beta \right) \quad,\quad \Be_{\Bu} \left(\Bxi,\Beta \right) = \frac{\Beta + \Bxi}{ \left\vert \Beta+\Bxi \right\vert }
	\label{eq_bondforce}
\end{equation}
Here $c(\Bxi)$ is the so-called bond micro-modulus which for an isotropic bulk material depends on the bond length $\xi = \left\vert \Bxi \right\vert$ only. The bond energy density can then be written in terms of the bond length $\xi$ and bond stretch $s$ as
\begin{equation}
	w \left(\Bx,\Bx^{\prime}\right) = 	w \left(\Bx^{\prime},\Bx\right) = \frac{c(\xi)}{2} s^2 \xi.
	\label{eq_bondenergy}
\end{equation}
The total elastic energy of the body ${\cal B}$ is obtained by integrating over all pairs of interacting material points: 
\begin{equation}
	E = \frac{1}{2} \int_{\cal B} \int_{{\cal H}_{\Bx}} w \left(\Bx,\Bx^{\prime}\right) \diff \Bx^{\prime} \diff \Bx = \int_{\cal B} W_{\textrm{p}}(\Bx) \diff \Bx .
	\label{eq_totalenergy}
\end{equation}
Note that the factor $\frac{1}{2}$ is needed in order not to count bonds twice in the double integration. In Eq. (\ref{eq_totalenergy}), we have for comparison with classical and gradient elasticity theories introduced the strain energy density $W_{\textrm{p}}(\Bx)$ associated with the point $\Bx$, defined as  
\begin{equation}
	W_{\textrm{p}} = \frac{1}{2} \int_{{\cal H}_{\Bx}} \!\! w \left(\Bx,\Bx^{\prime}\right) \diff \Bx^{\prime}.
	\label{eq:strainenergydensity}
\end{equation}
For the purpose of comparing peridynamics with classical or gradient elasticity theories, it is convenient to express the energy density in the limit of small deformations, $|\Beta| \ll |\Bxi|$, to 
obtain a force balance equation that is linear in the displacement field $\Bu$, and an energy functional that is a quadratic form of $\Bu$ \cite{silling2010peridynamic}.  This gives
\begin{equation}
	\Bf(\Bx,\Bx') = \BC(\Bxi)[\Bu(\Bx')-\Bu(\Bx)].
	\label{eq_bondforcelin}
\end{equation}
where 
\begin{equation}
	\BC(\Bxi)= \frac{c(\xi)}{\xi}[\Be_{\Bxi}\otimes \Be_{\Bxi}]
	\label{eq_micromod}
\end{equation}
and $\Be_{\Bxi} = \Bxi/\xi$. The strain energy density $W_{\textrm{p}}(\Bx)$ associated with the point $\Bx$ is then written as
\begin{equation}
	W_{\textrm{p}}(\Bx) = \frac{1}{4} \int_{{\cal H}_{\Bx}} \frac{c(\xi)}{\xi}  [\Be_{\Bxi} \cdot [\Bu(\Bx')-\Bu(\Bx)]]^2\diff \Bx'.
	\label{eq_strainenergydensity2}
\end{equation}
For comparison with classical or gradient elasticity theories, it is often useful to compute stress fields. Stresses, in classical continuum mechanics often defined as the work conjugates of strain variables, are however not a natural concept in peridynamics which is based upon displacements and not strains. Therefore, we use the analogy with particle systems where stresses are normally computed on a particle level using the virial theorem. Thus we evaluate the virial stress, which we take as a proxy of the Cauchy stress, as
\begin{equation}
	\Bsigma(\Bx) = \frac{1}{2} \int_{{\cal H}_{\Bx}} \Beta(\Bx,\Bx')\otimes \Bf(\Bx,\Bx')  \diff \Bx'.
	\label{eq_virialstress}
\end{equation}

\subsection{Relation with gradient elasticity}\label{sec_gradela}

To establish the relationship between bond-based peridynamics and higher-order strain gradient elasticity, we expand the displacement $\Bu$ into a Taylor series around the point $\Bx$. We define the first, second and third-order displacement gradients via
\begin{eqnarray}
	\nabla \Bu =: f \quad&,&\quad f_{ij} = u_{j,i} \nonumber\\
	\nabla \nabla \Bu = f' \quad&,&\quad f'_{ijk} = u_{k,ij} \nonumber\\
	\nabla \nabla \nabla \Bu = f'' \quad&,&\quad f''_{ijkl} = u_{l,ijk}	
\end{eqnarray}
where we use the notation $u_{j,i} := \partial u_j/\partial x_i$ and similarly for higher-order derivatives. The expansion of the displacement then reads
\begin{equation}
	u_i(\Bx') = u_i(\Bx) + f_{ji}(\Bx) \xi_j + \frac{1}{2} f'_{kji} \xi_k \xi_j + \frac{1}{6} f''_{jkli} \xi_j \xi_k \xi_l \dots
\end{equation}
where we use the standard summation convention. 
Upon insertion in \autoref{eq_strainenergydensity2}, we obtain an expansion of the energy density in terms of higher-order displacement gradients:
\begin{eqnarray}
	W_{\rm p}(\Bx) &=&  W_{\rm p}^{(0)} + W_{\rm p}^{(1)} + \dots ,\\
	W_{\rm p}^{(0)} &=&
	\frac{1}{4} \int_{{\cal H}_{\Bx}} \frac{c(\xi)}{\xi^3}f_{ij}(\Bx) f_{kl}(\Bx) \xi_i\xi_j\xi_k \xi_l  \diff\Bx',\\
	W_{\rm p}^{(1)} &=&
	\frac{1}{16} \int_{{\cal H}_{\Bx}} \frac{c(\xi)}{\xi^3}f'_{ijk}(\Bx) f'_{lmn}(\Bx) \xi_i\xi_j\xi_k \xi_l \xi_m \xi_n  \diff\Bx'\\
	\label{eq:strainEnergyDensity3}
\end{eqnarray}
We separate radial and angular parts of the integrals by writing $\xi_i = \xi e_{i}$ where the $e_i$ are components of a unit vector in $\Bxi$ direction and depend only on angular coordinates:
\begin{eqnarray}
	W_{\rm p}^{(0)} &=& f_{ij}(\Bx) f_{kl}(\Bx)
	\frac{1}{4} \int_{0}^{\delta_{\textrm{p}}} c(\xi)\xi^{D} \diff \xi \int_{\Omega^D}
    e_i e_j e_k e_l  \diff\Omega,\\
	W_{\rm p}^{(1)} &=& f'_{ijk}(\Bx) f'_{lmn}(\Bx)
	\frac{1}{16} \int_{0}^{\delta_{\textrm{p}}} c(\xi)\xi^{D+2} \diff \xi \int_{\Omega^D}
	e_i e_j e_k e_l e_m e_n  \diff\Omega,
	\label{eq:strainEnergyDensity2}
\end{eqnarray}
where $\Omega^D$ is the solid angle in $D$ dimensions and $\diff \Omega$ the corresponding angle element. Now it is easy to see that the angular integrations give zero unless the indices of the $e_i$ factors are pair-wise equal. At lowest order we find:
\begin{equation}
W_{\rm p}^{(0)} = \frac{1}{2} C_{ijkl} f_{ij}(\Bx) f_{kl}(\Bx) 
\end{equation}
where 
\begin{equation}
C_{ijkl} = \frac{\Omega_{\rm D}}{2} \int_{0}^{\delta_{\textrm{p}}} c(\xi)\xi^{D} \diff \xi 
[\delta_{ij} \delta_{kl} + \delta_{ik} \delta_{jl} + \delta_{il} \delta_{jk}] .
\end{equation}
For comparison with strain-based elasticity theory formulations it is convenient to split the deformation gradient into symmetric and antisymmetric parts, $f_{ij}  = \epsilon_{ij} + \omega_{ij}$. The antisymmetric part of the first-order deformation gradient does not contribute to the energy, and we write the lowest-order elastic energy contribution as
\begin{equation}
	W_{\rm p}^{(0)} = \frac{1}{2} \mu \left[  \epsilon_{ii}\epsilon_{kk} + 2 \epsilon_{ik}\epsilon_{ik} \right]
\end{equation}
which is the elastic energy of a classical linear-elastic material with Lam\'e constants
\begin{equation}
\lambda = \Omega_{\rm D} \int_{0}^{\delta_{\textrm{p}}} c(\xi)\xi^{D} \diff \xi
\end{equation}
and $\mu = \lambda$, hence Poisson number $\nu = 1/4$. 

We turn to the first-order contribution, following a similar line of argument. We write
\begin{equation}
	W_{\rm p}^{(1)} = \frac{1}{2} D_{ijklmn} f'_{ijk}(\Bx) f'_{lmn}(\Bx) 
\end{equation}
and introduce the characteristic length $l$ via
\begin{equation}
	\int_{0}^{\delta_{\textrm{p}}} c(\xi)\xi^{D+2} \diff \xi := l^2 \int_{0}^{\delta_{\textrm{p}}} c(\xi)\xi^{D} \diff \xi.
 \label{eq:lsquare}
\end{equation}
The first-order coupling constants can then be written as
\begin{eqnarray}
	D_{ijklmn} = \frac{\lambda l^2}{8} & \{ &
	[\delta_{ij} \delta_{kl} \delta_{mn} + \delta_{ij} \delta_{km}\delta_{ln} + 
	\delta_{ik} \delta_{jn} \delta_{lm} + \delta_{in} \delta_{jk}\delta_{lm}] \nonumber\\
	&+&
	[\delta_{ik} \delta_{jl} \delta_{mn} + \delta_{ik} \delta_{jm}\delta_{ln} + 
	\delta_{il} \delta_{jk} \delta_{mn} + \delta_{im} \delta_{jk}\delta_{ln}] \nonumber\\
    &+&
    \delta_{ij} \delta_{kn} \delta_{lm}\nonumber\\
    &+&    
    [\delta_{il} \delta_{jm} \delta_{kn} + \delta_{im} \delta_{jl}\delta_{kn}]
    \nonumber\\
    &+&
    [\delta_{il} \delta_{jn} \delta_{km} + \delta_{in} \delta_{jl}\delta_{km} + 
    \delta_{im} \delta_{jn} \delta_{kl} + \delta_{in} \delta_{jm}\delta_{kl}]
    \},
\end{eqnarray}
where we have grouped terms that, after summation over equal indices, yield equal energy contributions. Using the symmetry $f'_{ijk} = f'_{jik}$ and re-labeling we can write the gradient contribution to the elastic energy as
\begin{equation}
	W_{\rm p}^{(1)} = \frac{\lambda l^2}{16} \left[4 f'_{iik} f'_{kjj} + 4 f'_{ijj}f'_{ikk} 
	+  f'_{iik}f'_{jjk} + 2 f'_{ijk}f'_{ijk} + 4 f'_{ijk}f'_{ikj}\right] 
\end{equation}
Again we may perform a transition to strain gradients. Defining the strain gradient via $\eta_{ijk} := \epsilon_{jk,i}$, we obtain 
\begin{equation}
	W_{\rm p}^{(1)} = \frac{\lambda l^2}{16} \left[4 \eta_{iik} \eta_{kjj} + \eta_{ijj}\eta_{ikk} + 4 \eta_{iik}\eta_{jjk} 
	 + 2 \eta_{ijk}\eta_{ijk} + 4 \eta_{ijk}\eta_{kji}\right] .
\end{equation}
Either way, we find that up to first-order strain gradients the energy functional corresponds to a special case of Mindlin's energy functional for isotropic first gradient elasticity \cite{mindlin1968first}, where the five strain gradient coefficients $a_1 \dots a_5$ are given by
\begin{equation}
	a_1 = a_3 = a_5 = \frac{\lambda l^2}{4} \;,\; a_2 = \frac{\lambda l^2}{16} \;,\; a_4 = \frac{\lambda l^2}{8}.
    \label{eq:mindlincoeff}
\end{equation}
For standard parameterizations of the micro-modulus functions, the length $l$ is proportional to the horizon radius $\delta_{\textrm{p}}$, e.g. in 2D, for a constant micro-modulus $c(\xi) = c_0$, we obtain $l^2 = \frac{3}{5}\delta_{\textrm{p}}^2$, and for a 'conical' micro-modulus function $c(\xi) = c_0[1-\xi/\delta_{\textrm{p}}]$, $l^2 = \frac{2}{5}\delta_{\textrm{p}}^2$. 

\section{Constructing dislocations in peridynamics}
\label{sec:disloc}

\subsection{General formalism}\label{sec_general_formalism}

\begin{figure*}[htb]
	\includegraphics{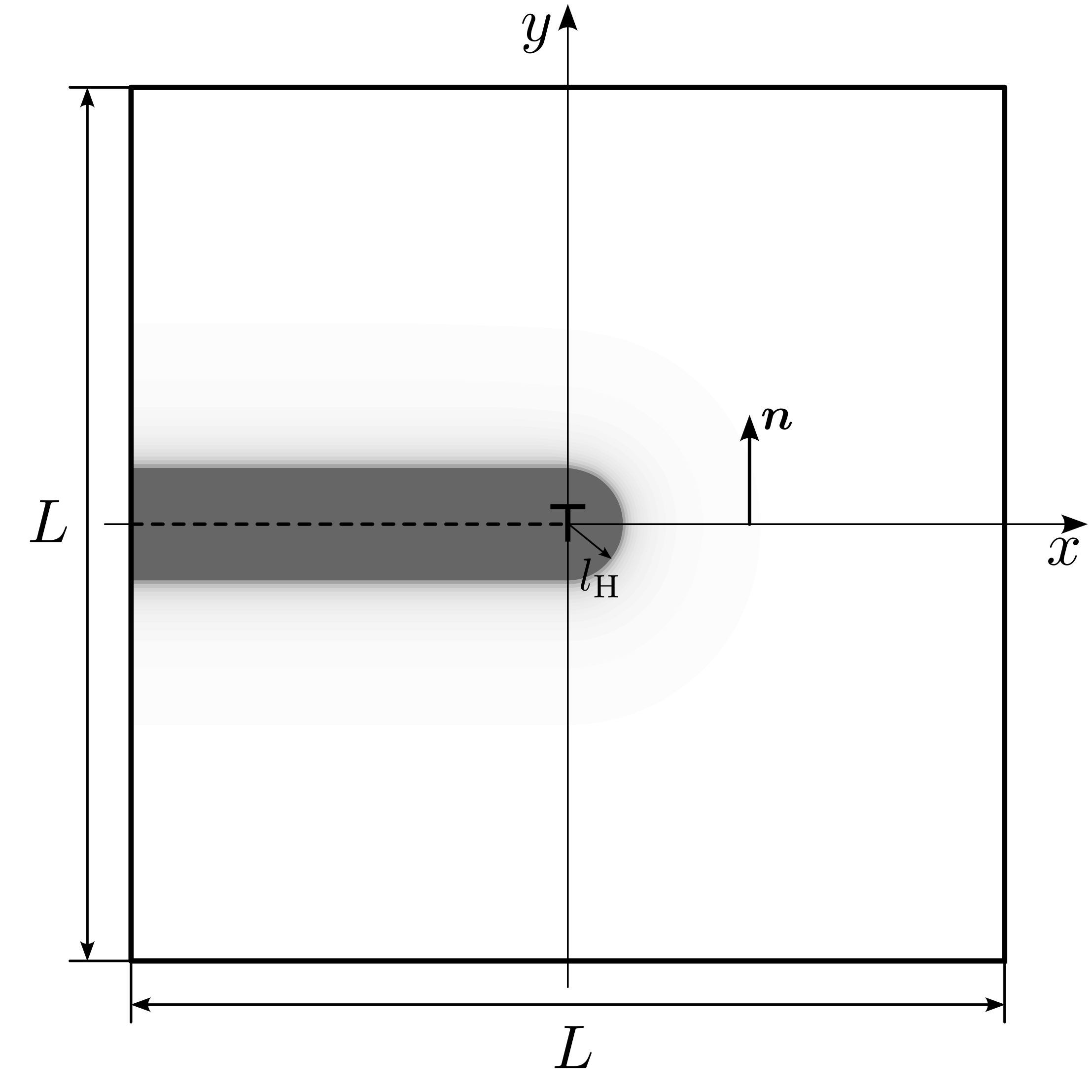}
	\caption{Schematic of a single edge dislocation with slip plane normal vector $\bm{n}$. The origin of the coordinates is centred in the considered domain and the grey area indicates the domain where the plastic distortion is provided by the gradient theory solution; to match peridynamics and Helmholtz type gradient elasticity, the horizon $\delta_{\textrm{p}}$ must be taken proportional to the length scale parameter $l_{\textrm{H}}$ of the gradient theory.}
	\label{fig:schematic}
\end{figure*}

The approach to constructing dislocations in peridynamics builds upon concepts from the classical continuum theory of dislocations. This theory relates dislocations to gradients of the plastic distortion, which in turn is envisaged in terms of slip on crystallographic slip planes. In the following a single dislocation line is considered as schematically illustrated in Fig. \ref{fig:schematic} for an edge dislocation.
We take the slip plane without loss of generality to be the plane $y=0$ with normal vector $\Bn = \Be_y$, and the Burgers vector as $\Bb = b \Be_x$. The dislocation line is parameterized as $\Br_{\textrm{d}}(s)$, where $\Br_{\textrm{d}} \cdot \Bn = 0 \;\forall\; s$ and $s$ is the line length. The unit tangent vector to the dislocation line is denoted by $\Bt(s) = \partial_s \Br_{\textrm{d}}$ and $\Bg = \Bt \times \Bn$ is the dislocation glide vector.

In the spirit of a continuum theory, we allow for the dislocation to be a continuously distributed object, by considering the Burgers vector to be distributed around the dislocation line. To describe this distribution, we use a scalar function $\phi(\Br(s),\Br_{\perp})$ where $\Br_{\perp} \!\!\cdot \Bt(s) = 0$, i.e., $\Br_{\perp}$ is the position vector in the plane locally perpendicular to the dislocation line, with the dislocation line located at $\Br_{\perp} = 0$. The function $\phi$ is normalized, i.e. $\int \phi \diff \Br_{\perp} = 1$. The corresponding dislocation density tensor field is given by
\begin{equation}
	\Balpha(\Br) =   [\Bt \otimes \Bb] \phi(\Br(s),\Br_{\perp})
\end{equation}
This tensor is linked to the plastic distortion $\Bbetapl$ via
\begin{equation}
	\Balpha(\Br) =   \Bnabla \times \Bbetapl. 
\end{equation}
So far, these are general relations used in the continuum theory of dislocations. The connection with 
peridynamics is established by noting that the plastic distortion represents a stress-free deformation which, in bond-based peridynamics, can be understood in terms of the force-free deformation of bonds. We parameterize a bond $\Bxi$ connecting to $\Bx$ as $\Br(\Bxi,\Bx,s) = \Bx + \Be_{\xi} s,\; 0 \le s \le \xi$.  The bond vector in the plastically deformed configuration is then evaluated as
\begin{equation}
	\Bxi' = \Bxi +  \int_0^{\xi} \Bbetapl \left( \Br(s) \right) \cdot \Be_{\xi} \, \diff s .
	\label{bonddef}
\end{equation}
with length $\xi' = \left\vert \Bxi' \right\vert$. The plastic bond deformation does not create a force but simply changes the bond vector and bond vector length entering the force calculation, which still is conducted according to Eq. (\ref{eq_bondforce}) or (\ref{eq_bondforcelin}).  

Once the modified bond vectors are computed, one more step is needed which amounts to a re-definition of the interaction sphere (`horizon'). This needs now to be defined in the bond-deformed configuration: 
\begin{equation}
	{\cal H}'_{\Bx} = \left\{\Bx'\left\vert \xi'(\Bx,\Bx') \le \delta_{\textrm{p}}\right. \right\}.
	\label{horizon2}
\end{equation}
In other words, the plastic bond deformation will convect some material points out of the original sphere of influence, while some new ones enter. This is in line with the idea that plastic deformation implies a change of atomic neighborhood relations and atomic interactions, while the physical nature of the environment of an atom remains unchanged (here: the influence region remains spherical, hence material isotropy is preserved if deformation is homogeneous). 

\subsection{Singular edge dislocation}

To illustrate the above ideas, we first consider the case of a singular dislocation which induces a discontinuity of slip along its glide plane. We consider a straight edge dislocation running along the $z$-axis with Burgers vector $\Bb = b \Be_x$, defining a plane strain problem in the perpendicular plane. The corresponding density function is $\phi(\Br_{\perp}) = \delta(\Br_{\perp})$, where here $\delta ( \cdot )$ is the Dirac delta distribution. In the following, the subscript $\perp$ is dropped since a 2D, plane strain implementation in peridynamics is envisaged, where only space dependence in the $xy$-plane needs to be considered. The dislocation tangent vector is $\Bt = \Be_z$, and the slip plane normal vector is $\Bn = \Be_y$. The singular dislocation density tensor follows as
\begin{equation}
	\Balpha(\Br) = b \left[ \Be_z \otimes \Be_x \right] \phi(\Br)
\end{equation}
and the plastic distortion field  which fulfills $\Balpha = \Bnabla \times \Bbetapl$ is, up to a constant offset, given by 
\begin{equation}
	\Bbetapl(x,y) = b \left[ \Be_x \otimes \Be_y \right] \delta(y) H(-x) = \frac{b}{\delta_{\textrm{p}}}  \left[ \Be_x \otimes \Be_y \right] \delta \left(\frac{y}{\delta_{\textrm{p}}}\right)H\left(-\frac{x}{\delta_{\textrm{p}}}\right)
	\label{betaplsing}
\end{equation}
where $H$ is Heaviside's unit step function and the second, equivalent term on the right-hand side has been added for later use.  From the non analytic function given by Eq. (\ref{betaplsing}), the bond displacement is calculated via Eq. (\ref{bonddef}). The result can be stated in simple words: If the bond vector crosses the slipped parts of the slip plane (i.e. in 2D, the negative $x$-axis) in $+y$-direction, we add $\Bb$, if it crosses the slip plane in $-y$-direction, we subtract $\Bb$.

\subsection{Edge dislocation with distributed Burgers vector}

\begin{figure*}[h]
	\includegraphics[width=\textwidth]{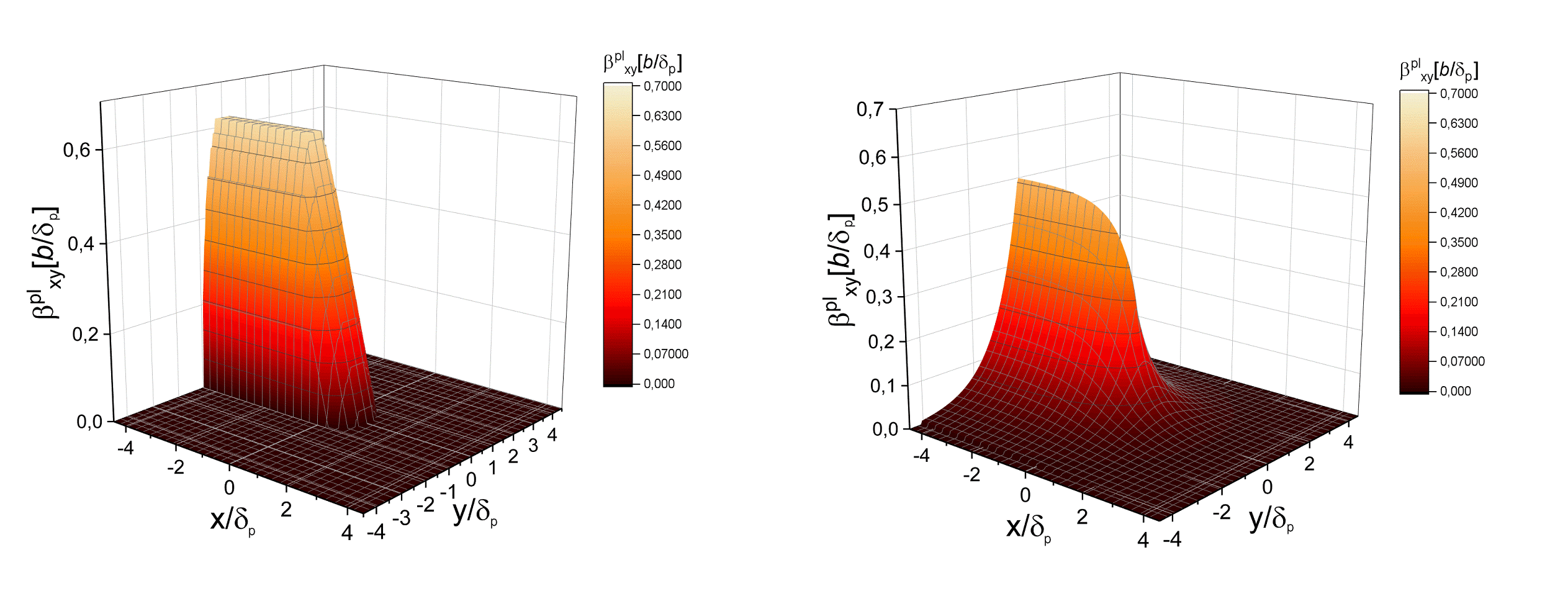}
	\caption{Plastic distortion fields of distributed dislocations in peridynamics, Burgers vector points in $x$-direction and the glide plane trace corresponds to the $x$-axis, left graph: constant Burgers vector distribution given by Eq. (\ref{burgers1}); right graph: Burgers vector distribution given by Eq. (\ref{burgers2}).}
	\label{fig:betaxy}
\end{figure*}

The concept of a singular point-like dislocation sits uneasily with the general spirit of peridynamics, where interactions are considered of finite range. A straightforward generalization is to consider a Burgers vector distribution $\phi$ of finite range. For illustration, we consider the same edge dislocation as in the previous section but now assume that the Burgers vector is evenly distributed over a circle of radius $\delta_{\textrm{p}}$ around the origin:
\begin{equation}
	\displaystyle{
		\phi(\Br) = \left\{\begin{array}{ll} \displaystyle\frac{1}{\pi\delta_{\textrm{p}}^2},& \displaystyle\frac{r}{\delta_{\textrm{p}}} < 1\\[12pt]
			0, & {\textrm{otherwise,}}\end{array}\right.}
	\label{burgers1}
\end{equation}
where $r = \left\vert \bm{r} \right\vert$. Alternatively, we consider a Burgers vector distribution which is 'designed' to match the field of the dislocation in gradient elasticity. To this end, the Green's function of the Helmholtz equation is used as Burgers vector distribution:
\begin{equation}
	\phi(\Br) = \frac{1}{2\pi l_{\textrm{H}}^2} K_0\left(\frac{r}{l_{\textrm{H}}}\right)
	\label{burgers2}
\end{equation}
with $l_{\textrm{H}} = \delta_{\textrm{p}} \sqrt{3/5}$.
The distributed Burgers vector induces a distributed dislocation density tensor and a spatially distributed plastic distortion which is not explicitly given but computed numerically, with results shown in \figref{fig:betaxy}. Again, the bond deformation follows by using Eq. (\ref{bonddef}). 

\subsection{Scaling property of the dislocation related fields}

We now assume that the dislocation related fields can be computed using a small strain approximation where $b \ll \delta_{\textrm{p}}$. If, moreover, the same length scale $\delta_{\textrm{p}}$ governs the range of elastic interactions and the non-local distribution of strain, as in our above example, then the ensuing fields exhibit scaling properties which are well-known in the context of classical dislocation theory \cite{zaiser2014scaling,berdichevsky2023temperature}. In particular, the stress field $\Bsigma$, plastic distortion $\Bbeta^{\textrm{pl}}$ and elastic distortion $\Bbeta^{\textrm{el}}$ all have the scaling form 
\begin{equation}
	\BX(\Br)  = \frac{b}{\delta_{\textrm{p}}}\tilde{\BX}\left(\frac{\Br}{\delta_{\textrm{p}}}\right)\quad,\quad \BX \in \{\Bsigma,\Bbeta^{\textrm{pl}},\Bbeta^{\textrm{el}}\}
	\label{scaleX}
\end{equation}
and the displacement field has the form
\begin{equation}
	\Bu(\Br)  = b\tilde{\Bu}\left(\frac{\Br}{\delta_{\textrm{p}}}\right) ,
	\label{uscale}
\end{equation}
Eq. (\ref{scaleX}) holds for the singular distortion field given by Eq. (\ref{betaplsing}) as demonstrated by the second term on the right hand side of that equation. Accordingly, the plastic bond displacements computed from Eq. (\ref{bonddef}) obey Eq. (\ref{uscale}). Moreover, the elastic displacements that arise in response to the bond deformation must be proportional to the same owing to the linearity of Eq. (\ref{eq_bondforcelin}). The scaling property for the stress field then follows by noting that, in the small strain limit, $\eta \approx \xi \approx \xi'$ is used in the calculation of the virial stress via Eq. (\ref{eq_virialstress}).  It is noted that the stress, distortion and displacement fields of a dislocation in classical elasticity fulfill Eqs. (\ref{scaleX}) and (\ref{uscale}) for arbitrary $\delta_{\textrm{p}}$. Finally it is noted that, in the small strain limit and in an infinite body, the field of a distributed dislocation follows from that of a singular distribution by convolution with the Burgers vector distribution provided this field again has the above formulated scaling form.

\subsection{Dislocations in gradient elasticity theory}

The elasticity theory of dislocations in Mindlin's first gradient elasticity has been discussed extensively by Lazar \cite{lazar2021incompatible}, considering both the general theory \cite{lazar2021incompatible} and simplified versions with a single length scale ('Helmholtz-type gradient elasticity' \cite{lazar2005nonsingular,lazar2013fundamentals}). Here, we focus on the case of edge dislocations which represent a plane strain problem. At first, the main results of the general theory, as applied to edge dislocations by Lazar \cite{lazar2021incompatible}, are summarized. 

In the full theory, the stress field of an edge dislocation is controlled by four length scales which relate to the gradient coefficients $a_1\dots a_5$, Eq. (\ref{eq:mindlincoeff}), via \cite{lazar2021incompatible}
\begin{equation}
	\begin{aligned}
		l_1^2 &= \frac{2(a_1+a_2+a_3+a_4+a_5)}{\lambda + 2 \mu},\\
		l_2^2 &= \frac{a_3+2a_4+a_5}{2\mu},\\
		l_3^2 &= \frac{2(2a_1+3a_2+a_3+a_4+a_5)}{3\lambda + 2 \mu},\\
		l_4^2 &= \frac{a_1+2a_3+2a_4+2a_5)}{2 \mu}.\\
		\label{mindlinlengths}
	\end{aligned}
\end{equation}
The displacement, elastic and plastic distortion fields as well as the stress field of the dislocation can be expressed in terms of these four characteristic lengths in conjunction with the elastic constants of the material and the dislocation Burgers vector. Explicit expression for the stress field are given by Lazar \cite{lazar2021incompatible}, considering an edge dislocation with Burgers vector $\Bb = b \Be_x$: 
\begin{equation}
	\begin{aligned}
		\sigma_{xx} \,=\, &- \frac{\mu b}{2\pi(1-\nu)}\frac{y}{r^2}\left\{\frac{3x^2+y^2}{r^2}-\frac{2 \nu r}{l_1}K_1(r/l_1) \right.\\
		&\left.
		-2(1-\nu)\left[\frac{3x^2-y^2}{r^2}\left(\frac{4l_2^2}{r^2}-2K_2(r/l_2)\right) - \frac{x^2-y^2}{l_2r}K_1(r/l_2)\right]\right.\\
		&\left.
		+(1-2\nu)\left[\frac{3x^2-y^2}{r^2}\left(\frac{4l_1^2}{r^2}-2K_2(r/l_1)\right) - \frac{2x^2}{l_1r}K_1(r/l_1)\right]
		\right\}\\
		\sigma_{yy} \,=\, &\frac{\mu b}{2\pi(1-\nu)}\frac{y}{r^2}\left\{\frac{x^2-y^2}{r^2}+\frac{2 \nu r}{l_1}K_1(r/l_1)\right.\\
		&\left.
		-2(1-\nu)\left[\frac{3x^2-y^2}{r^2}\left(\frac{4l_2^2}{r^2}-2K_1(r/l_2)\right) - \frac{x^2-y^2}{l_2r}K_1(r/l_2)\right]\right\}\\
		&\left.
		+(1-2\nu)\left[\frac{3x^2-y^2}{r^2}\left(\frac{4l_1^2}{r^2}-2K_2(r/l_1)\right) + \frac{2y^2}{l_1r}K_1(r/l_1)\right]\right\}\\
		\sigma_{xy} \,=\, &\frac{\mu b}{2\pi(1-\nu)}\frac{x}{r^2}\left\{\frac{x^2-y^2}{r^2}+2(1-\nu)\left[\frac{r}{l_2}K_1(r/l_2)
		-\frac{r}{l_4}K_1(r/l_4)\right]\right.\\
		&\left.
		-2(1-\nu)\left[\frac{x^2-3y^2}{r^2}\left(\frac{4l_2^2}{r^2}-2K_2(r/l_2)\right) + \frac{2y^2}{l_2r}K_1(r/l_2)\right]\right.\\
		&\left.
		+(1-2\nu)\left[\frac{x^2-3y^2}{r^2}\left(\frac{4l_1^2}{r^2}-2K_2(r/l_1)\right) + \frac{2y^2}{l_1r}K_1(r/l_1)\right]\right\}\\
		\sigma_{zz} = &- \nu (\sigma_{xx}+\sigma_{yy}) = - \frac{\mu b \nu}{\pi(1-\nu)}\frac{y}{r^2}\left\{1-\frac{r}{l_1}K_1(r/l_1)\right\}.
		\label{mindlinstress}
	\end{aligned}
\end{equation}
Here the $K_i$ are modified Bessel functions of the second kind. It is noted that, as in the classical theory, the sum $\sigma_{xx} + \sigma_{yy}$ has the same functional form as the $yz$-stress field component of a screw dislocation. 

As discussed by Lazar \cite{lazar2021incompatible}, a simplified theory emerges if all four characteristic lengths $l_1\dots l_4$ are equal, $l_1 = l_2 = l_3 = l_4=l_{\textrm{H}}$, leading to so-called gradient elasticity of Helmholtz-type \cite{lazar2005nonsingular,lazar2013fundamentals}. This theory has the advantage that, just as in the classical theory of dislocations, the stress fields of general dislocation configurations can be expressed analytically in terms of integrals over the dislocation lines. For Helmholtz-type gradient elasticity, the relationships between the classical and gradient elasticity fields are exceedingly simple \cite{lazar2005nonsingular}:
\begin{equation}
	(1 - l_{\textrm{H}}^2 \Delta) \BX = \BX_0 \;,\; \BX \in \{\Bsigma,\Bbetapl,\Bbeta^{\textrm{el}},\Bu\},
	\label{scalehelmholtz}
\end{equation}
where $\Delta$ is the Laplace operator, $\BX$ denotes the field in gradient elasticity and $\BX_0$ its classical counterpart. In other words, the gradient elasticity fields can be obtained by convolution of the classical fields with the Green's function of the Helmholtz operator, 
\begin{equation}
	\phi\left(\frac{r}{l_{\textrm{H}}}\right) = \frac{1}{2\pi l_{\textrm{H}}^2} K_0\left(\frac{r}{l_{\textrm{H}}}\right)
\end{equation}
where $K_0$ is again a modified Bessel function of the second kind.

As to the scaling properties discussed in the previous section, we observe that the stresses given by Eq. (\ref{mindlinstress}) exhibit the same scaling properties as the peridynamic stress field. This can be seen by noting that, when the gradient elasticity is matched to peridynamics, according to Eqs. (\ref{eq:mindlincoeff}, \ref{mindlinlengths}), all four characteristic lengths of the gradient theory are proportional to $\delta_{\textrm{p}}$, and the scaling form of Eq. (\ref{scaleX}) is then, for the stress field, immediately evident from Eq. (\ref{mindlinstress}). As to the other fields, the scaling behavior of Eqs. (\ref{scaleX}) and (\ref{uscale}) immediately follows from Eq. (\ref{scalehelmholtz}) by noting that the Helmholtz operator is scaling invariant if $l_{\textrm{H}} \propto \delta_{\textrm{p}}$ and the classical fields exhibit the same scaling behavior as their peridynamic counterparts. 

\section{Simulation results and comparison with gradient elasticity}
\label{sec:peridis}

In the following, we consider a square system deforming in 2D plane strain. For discretization a regular square grid of 200 $\times$ 200 collocation points is used, and space is measured in units of the collocation point spacing. The origin of the coordinate system is set in the centre of the square which also defines the location of the dislocation, c.f. Fig. \ref{fig:schematic}. The horizon $\delta_{\textrm{p}}$ is varied between 3.01 and 12.01 units. The Burgers vector is taken to be $b = 0.1$ units. To be able to compute fields with meaningful units, 1 unit is taken to be \SI{1}{\nano \metre}, and a Young's modulus of \SI{1}{\giga \pascal} is assumed, noting that re-scaling for any desired material is straightforward. Unless otherwise stated, the micro-modulus is assumed to be constant over the influence sphere. 

\subsection{Singular edge dislocation}

\begin{figure*}[h]
	\includegraphics[width=\textwidth]{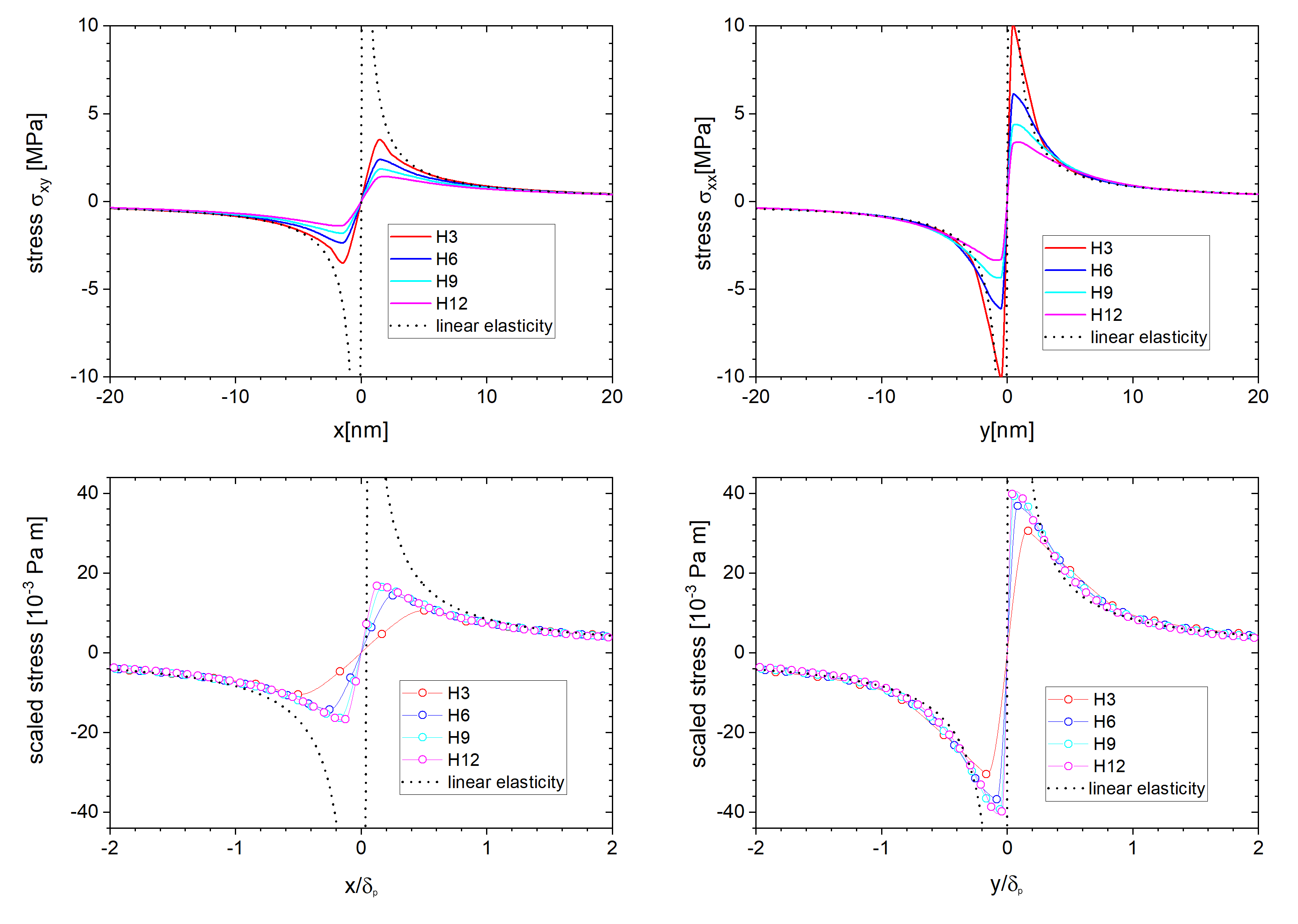}
	\caption{Stress field components of a singular dislocation in peridynamics, the Burgers vector points in $x$-direction and the glide plane trace corresponds to the $x$-axis; left: $\sigma_{xy}(x,y=0)$, right: $\sigma_{xx}(x=0,y)$; the top graphs show the original results for different values of the horizon where $\delta_{\textrm{p}} = \SIlist{3;6;9;12}{\nano \metre}$ (marked H3, 6, 9, 12 in the graph labels); the bottom graphs show the same data after rescaling $\Bsigma \to \delta_{\textrm{p}} \Bsigma, \Br \to \Br/\delta_{\textrm{p}}$; in these graphs the locations of the collocation points at which virial stresses are evaluated are marked by open circles.}
	\label{fig:singularscale}
\end{figure*}

We first consider the case of an edge dislocation with singular distribution of the plastic distortion, Eq. (\ref{betaplsing}). Resulting stress fields are shown in \figref{fig:singularscale} (top) for different values of the horizon $\delta_{\textrm{p}}$ as indicated in the graphs, which show sections of the $\sigma_{xy}$ shear stress component along the $x$ and of the $\sigma_{xx}$ component along the $y$-axis. It is noted that the functional shape of $\sigma_{yy}(x,y)$ is almost identical with $\sigma_{xy}(y,x)$, thus no separate plot is shown for this component.  

\begin{figure*}[htb]
	\includegraphics[width=\textwidth]{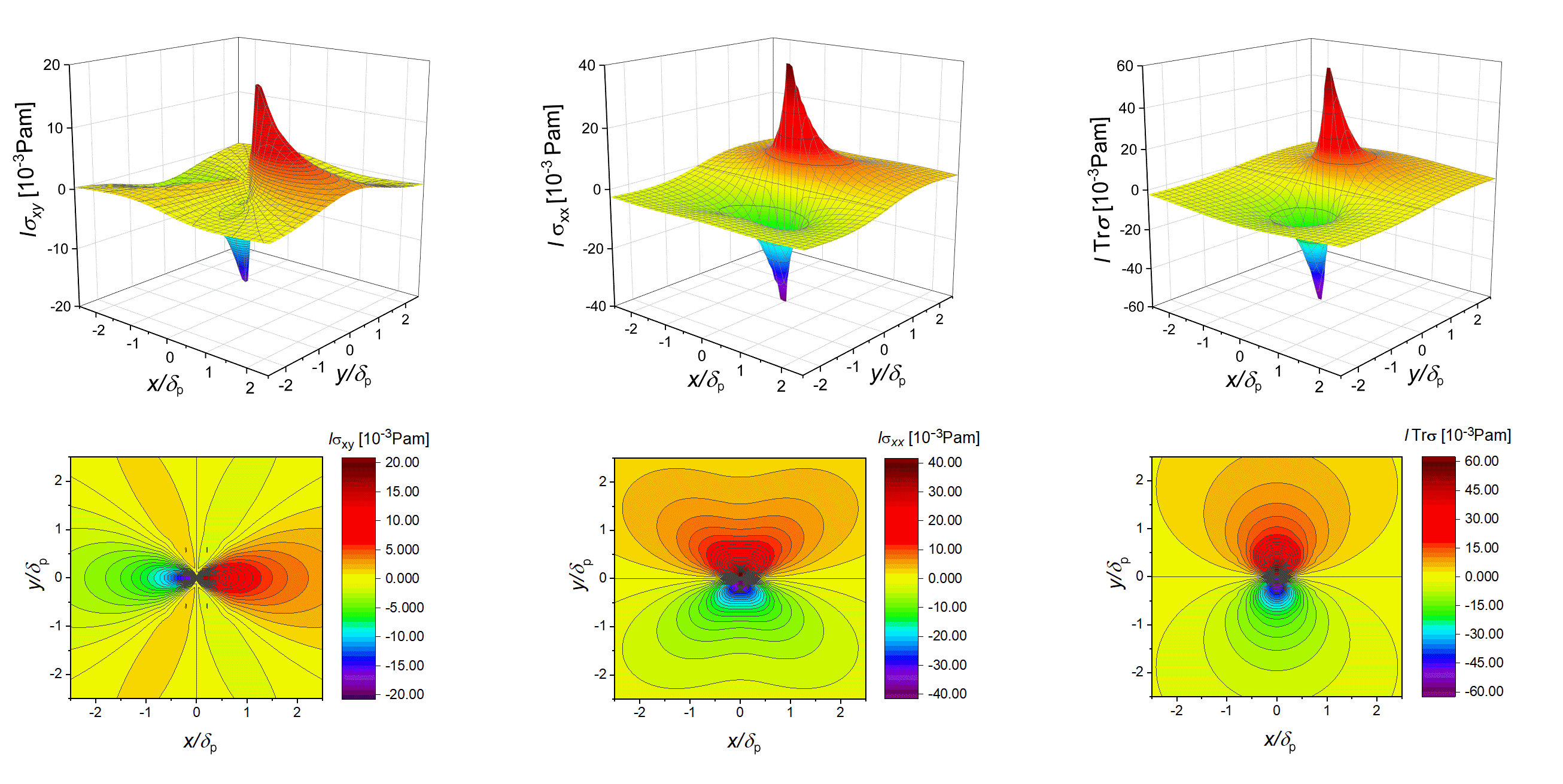}
	\caption{Angle dependence of the dislocation stress field components, $\delta_{\textrm{p}} = \SI{12}{\nano \metre}$, other parameters as in \figref{fig:singularscale}; top: 3D representation, bottom: contour plot.}
	\label{fig:anglescale}
\end{figure*}

\begin{figure*}[htb]
	\includegraphics[width=\textwidth]{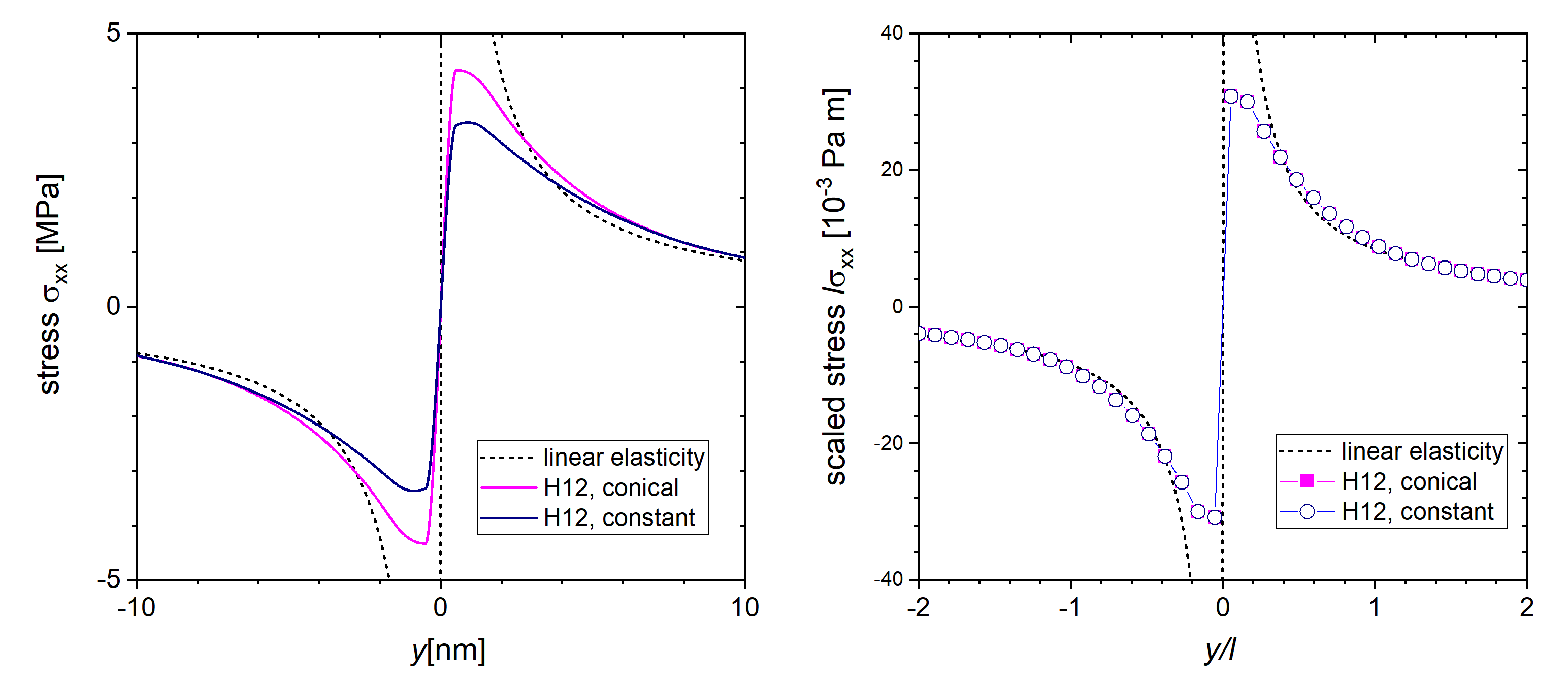}
	\caption{Effect of micro-modulus distribution on stress of a singular dislocation; all parameters as in \figref{fig:singularscale}; left: $\sigma_{xx}(x=0, y)$ for $\delta_{\textrm{p}} = \SI{12}{\nano \metre}$, simulated using a constant micro-modulus and a 'conical' micro-modulus distribution; right: scaling collapse obtained by plotting $l\sigma_{xx}$ vs. $y/l$ where $l = \delta_{\textrm{p}} \sqrt{3/5}$ for data evaluated with constant and $l= \delta_{\textrm{p}} \sqrt{2/5}$ for data evaluated with 'conical' micro-modulus.}
	\label{fig:horizonscale}
\end{figure*}

Different from the classical solution, the peridynamic stress fields do not exhibit a singularity in the origin. The deviation from the classical solution is evident at distances of the order of $\delta_{\textrm{p}}$ from the singularity (\figref{fig:singularscale}, top). At the same time, there is a kind of convergence to the classical solution as decreasing the horizon $\delta_{\textrm{p}}$ leads to an increase of the maximum stress near the dislocation and to a decrease of the distance over which differences between the peridynamic and classical solutions are evident. This qualitative observation was reported previously by Zhao et al. \cite{zhao2021nonlocal}, who however fail to realize the underlying mathematical scaling structure. This scaling structure is illustrated in \figref{fig:singularscale} (bottom), where components of the stress tensor, multiplied with $\delta_{\textrm{p}}$, vs. the re-scaled space coordinates $x/\delta_{\textrm{p}}$ and $y/\delta_{\textrm{p}}$ are plotted. According to Eq. (\ref{scaleX}), the result should be independent of $\delta_{\textrm{p}}$, which is approximately correct for $\delta_{\textrm{p}} \ge \SI{9}{\nano \metre}$; deviations at smaller values result from the fact that the number of collocation points within the horizon is too small. It is noted in this context, that the actual solution is difficult to represent because it is non analytic in the origin where both stress components $\sigma_{xx}$ and $\sigma_{xy}$ exhibit a jump, as can be seen from \figref{fig:singularscale}. Accordingly, a finite spacing of the collocation points used for stress evaluation may lead to the erroneous suggestion of smooth behavior in the origin, a problem that may be exacerbated by injudicious use of interpolation functions. 
The angular dependence of the stress fields matches their classical counterparts as illustrated in \figref{fig:anglescale}. 
We next analyze to which extent the observed behavior is contingent on the choice of the micro-modulus function. \figref{fig:horizonscale} compares results obtained for constant and 'conical' micro-modulus distribution, all other parameters being identical.
It can be seen in \figref{fig:horizonscale}, left, that the 'conical' distribution, which gives a higher weight to short bonds and leads to a smaller length parameter $l$, results in more pronounced positive and negative stress peaks located at the collocation points adjacent to the origin. However, if we exploit the scaling invariance expected for both gradient elasticity and peridynamics simulations, and plot $l \sigma_{xx}$ vs. $\Br/l$ where $l = \delta_{\textrm{p}} \sqrt{3/5}$ for constant and $l= \delta_{\textrm{p}} \sqrt{2/5}$ for 'conical' micro-modulus distribution, one finds a near-perfect collapse of both curves as seen in \figref{fig:horizonscale}, right. This indicates, on the one hand, that the non analytic behavior in the origin is not just an artefact of a particular choice of micro-modulus distribution, and demonstrates on the other hand that dislocation fields obtained from different distributions can be analyzed in a common mathematical framework which measures the range of non-local forces in terms of the length scale $l$ as given by Eq. (\ref{eq:lsquare}). 

\subsection{Distributed edge dislocation}

Next, the stress fields of dislocations evaluated with distributed Burgers vector are considered. All simulations reported in the following use a micro-modulus that is constant over the influence sphere. First, the case where the Burgers vector is evenly distributed over the influence sphere is considered, Eq. (\ref{burgers1}). 

\begin{figure*}[h]
	\includegraphics[width=\textwidth]{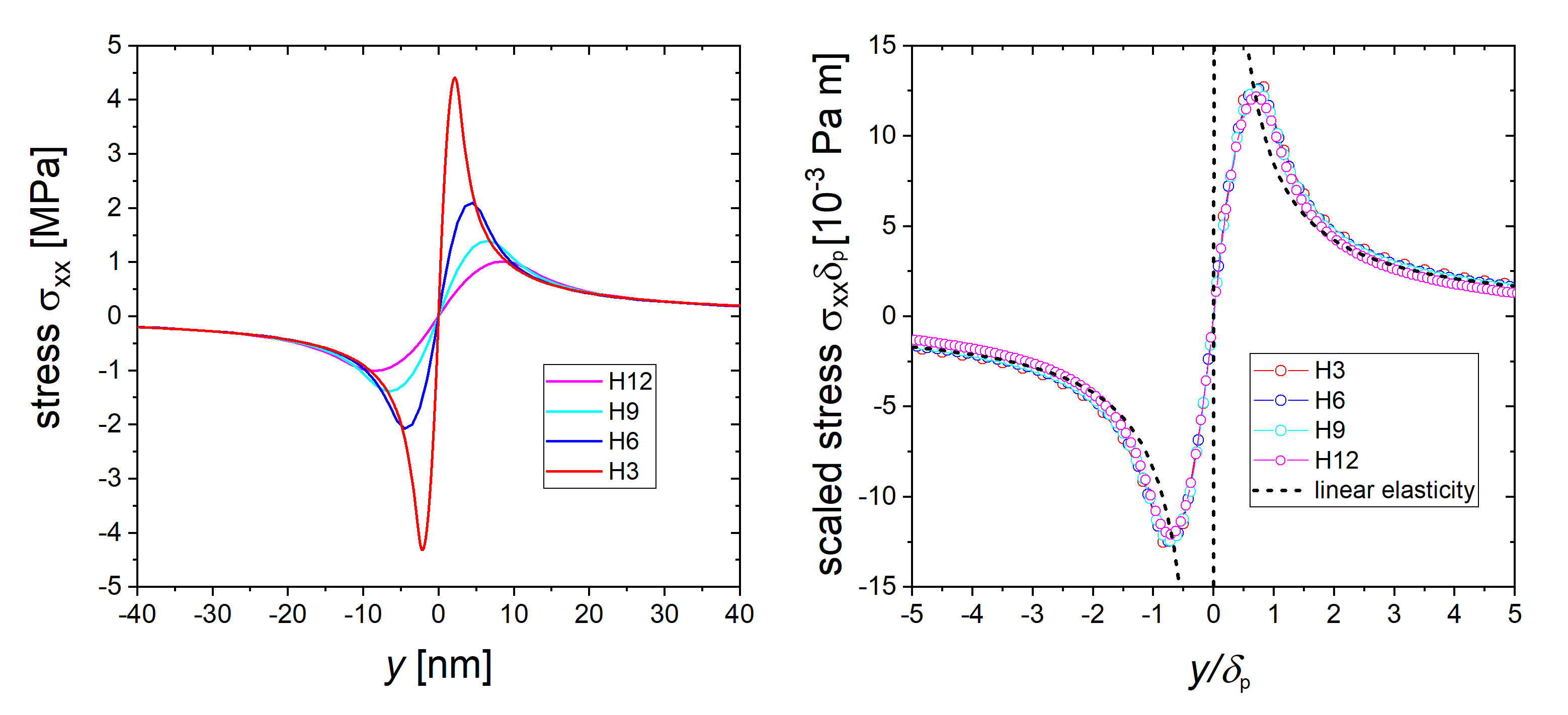}
	\caption{Stress field components of a distributed dislocation in peridynamics, the Burgers vector points in $x$-direction and the glide plane trace corresponds to the $x$-axis, Burgers vector distribution given by Eq. (\ref{burgers1}); left: $\sigma_{xy}(x = 0, y)$ for different values of the horizon, right: same data after rescaling $\Bsigma \to \delta_{\textrm{p}} \Bsigma, \Br \to \Br/\delta_{\textrm{p}}$; in these graphs the locations of the collocation points at which virial stresses are calculated are marked by open circles.}
	\label{fig:distributedscale}
\end{figure*}

\figref{fig:distributedscale} (left) gives profiles of the $\sigma_{xx}$ component along the $y$-axis, evaluated for different values of $\delta_{\textrm{p}} \in \{3, 6, 9, 12\} \, \si{\nano \metre}$.
Comparison with \figref{fig:singularscale} demonstrates that the peak stress values are reduced as compared to the singular dislocation, and that the non analytic behavior in the origin disappears. The scaling collapse in \figref{fig:distributedscale} (right) now indicates excellent superposition of the curves for all $\delta_{\textrm{p}}$ values, indicating much improved convergence of the dislocation fields towards the scaling solution, even for small numbers of collocation points within the horizon. This is important from the viewpoint of efficient numerical implementation since, for fixed physical values of the system size and the horizon, the computation time increases in proportion of the square of the number of collocation points within the horizon. 

\subsection{Comparison with dislocation in gradient elasticity}

\begin{figure*}[h]
	\includegraphics[width=\textwidth]{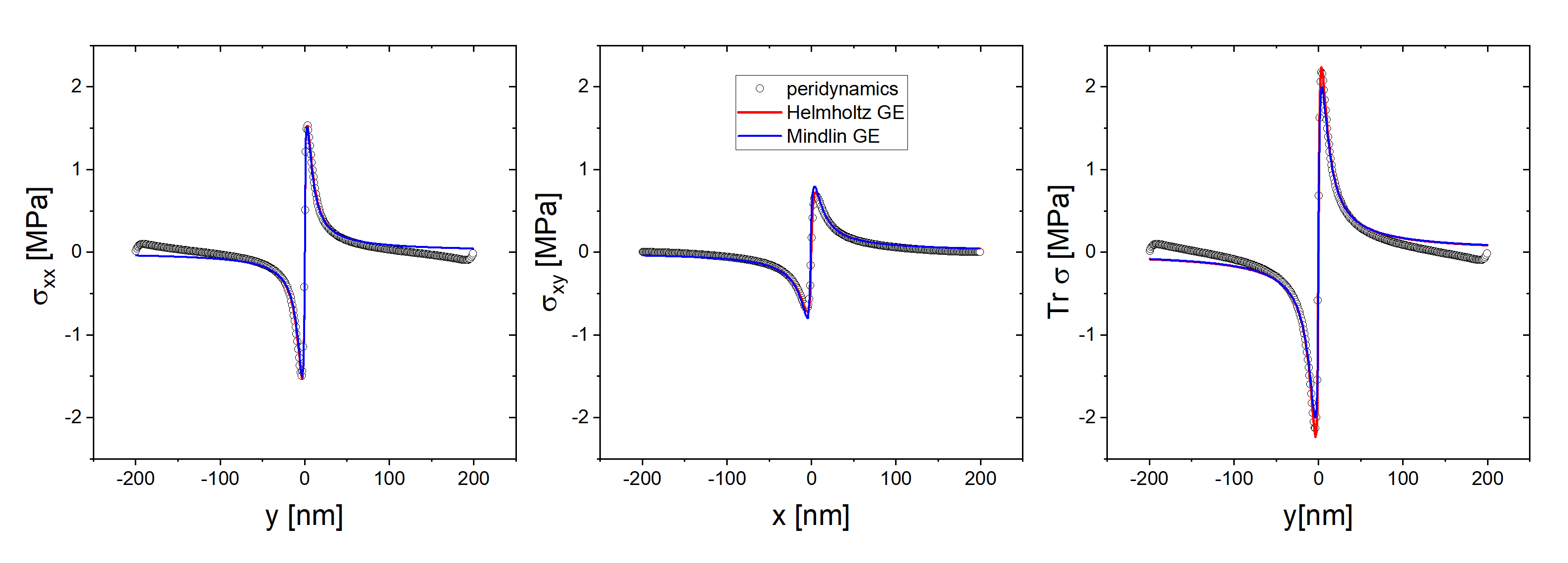}
	\caption{Stress field components of a distributed dislocation in peridynamics and comparison with gradient elasticity; Burgers vector points in $x$-direction and the glide plane trace corresponds to the $x$-axis, Burgers vector distribution given by Eq. (\ref{burgers2}) with horizon $\delta_{\textrm{p}} = \SI{3}{\nano \metre}$; left: $\sigma_{xx}(x=0,y)$, middle: $\sigma_{xy}(x,y=0)$, right: ${\rm Tr}\Bsigma(x=0,y)$; the open circles indicate virial stresses evaluated at the respective collocation points, the red line gives the matching solution evaluated using Helmholtz gradient elasticity, and the blue line the solution evaluated using (full) Mindlin gradient elasticity.}\label{fig:comparison}
\end{figure*}

To compare with gradient elasticity, a Burgers vector distribution is used that matches the one underlying the gradient elasticity solution, Eq. (\ref{burgers2}). Results obtained with this Burgers vector distribution are shown for $\delta_{\textrm{p}} = \SI{3}{\nano \metre}$ in \figref{fig:comparison}.

For comparison, two variants of gradient elasticity are shown. First, the solution of the full Mindlin equations as given by Eq. (\ref{mindlinstress}) is used, where the four length parameters $l_1\dots l_4$ were obtained from the value $\delta_{\textrm{p}} = \SI{3}{\nano \metre}$ using Eqs. (\ref{eq:mindlincoeff}, \ref{mindlinlengths}) in conjunction with $l = \delta_{\textrm{p}} \sqrt{3/5}$ for the constant micro-modulus distribution, that was used in the peridynamics calculations to obtain $l_1 = \SI{3.35}{\nano \metre}$, $l_2 = \SI{2.60}{\nano \metre}$, $l_3 = \SI{3.07}{\nano \metre}$, and $l_4 = \SI{3.67}{\nano \metre}$. This solution is represented by the blue lines in \figref{fig:comparison}. A simplified calculation is also performed, setting $l_1 = l_2 = l_3 = l_4 = l_{\rm H}=\SI{3}{\nano \metre}$ to obtain the Helmholtz gradient elasticity solution, which is shown in red in \figref{fig:comparison}. Both gradient elasticity solutions are quite similar, and both are in good agreement with the peridynamic solution in the central part of the simulated sample. Agreement is slightly better for the Helmholtz gradient elasticity solution, which is to be expected since the Burgers vector distribution has been chosen to match the Helmholtz gradient elasticity result. 

Different from previous figures, where we have focused on the region near the dislocation, \figref{fig:comparison} shows the stress tensor components for the entire domain from surface to surface. While the gradient elasticity results represent solutions in an infinite body which go to zero only asymptotically, the peridynamic solution automatically accounts for surface effects. Thus, near the surfaces, both gradient elasticity solutions differ from the peridynamic solution: On the upper and lower surfaces with normal vector $\Be_y$, the hydrostatic stress component $\sigma_{yy}$ is approximately zero for the peridynamic dislocation, as is the shear stress $\sigma_{xy}$ on the left and right surfaces with normal vector $\Be_x$. On the other hand, the hydrostatic stress component $\sigma_{xx}$ is non zero on the upper and lower surface. This behavior is in line with the surface boundary conditions imposed in classical elasticity theory on a free surface. The gradient elasticity stress fields, computed for a dislocation in an infinite body, assume finite values on all surfaces. 

\section{Construction of a hybrid multiscale dislocation model}
\label{sec:hybrid}

It has been shown that the fields associated with a dislocation in peridynamics have certain scaling properties, and that they can be matched with results for dislocations in gradient elasticity if an appropriate Burgers vector distribution is chosen. Moreover, the peridynamic dislocation satisfies surface boundary conditions for the stress fields while gradient elasticity solutions calculated for a bulk dislocation do not. The question arises what these findings are good for. 

First it is noted that peridynamics as such is not a tool of choice for dislocation simulations. If the horizon $\delta_{\textrm{p}}$ is taken much larger than the atomic spacing, as it should for computational reasons, then the stress fields close to the dislocation core are underestimated by a factor of the order of $b/\delta_{\textrm{p}}$ and accordingly the interactions of close dislocations are ill represented. These interactions are however crucial for our understanding of work hardening \cite{kubin2008modeling}; the problem is well known from attempts to describe dislocation interactions by FEM in so-called discrete-continuum schemes \cite{lemarchand2001homogenization} where the element size plays a similar role. If, on the other hand, the horizon is taken of the order of the atomic spacing, then the simulations become computationally very expensive and molecular dynamics is a both cheaper and more accurate option. However, the close match between numerical solutions obtained by peridynamics and analytical solutions obtained by gradient elasticity still allows us to devise a useful hybrid scheme. 

The method is based on using a combination of peridynamic and gradient elasticity for evaluating the stress field of a dislocation in a multiscale model as the sum of three contributions:
\begin{itemize}
	\item On the peridynamic side, a horizon $\delta_{\textrm{p}}$ is chosen which is computationally convenient and may be much larger than the physical extension of the dislocation core. With this horizon the methodology described above is used to construct a peridynamic dislocation with desired line direction and Burgers vector,  with the Burgers vector distribution given by Eq. (\ref{burgers2}). The resulting peridynamic stress field is denoted as $\Bsigma^{\textrm{p}}(\Br,\delta_{\textrm{p}})$. It is noted that, since this stress field is free of singularities or discontinuities near the dislocation core, it can be safely interpolated between the collocation points to evaluate stresses at any desired spatial resolution. 
	\item Using the relations of (Helmholtz) gradient elasticity, matching gradient elasticity fields with length scale parameter $l_{\textrm{H}} \sim \delta_{\textrm{p}}$ are evaluated. The corresponding stress field is denoted as $\Bsigma^{\textrm{g},\delta_{\textrm{p}}}(\Br)$. Near the dislocation core, it has the same behavior as the peridynamic stress field, whereas at distances $r \gg \delta_{\textrm{p}}$ from the dislocation it behaves like the classical bulk solution for the dislocation stress. 
	\item Again using Helmholtz gradient elasticity, a gradient elasticity field with a length scale parameter $l_{\textrm{b}} \sim b$ is evaluated. The parameter $l_{\textrm{b}}$ is chosen such to match the core properties of the dislocation on the atomistic level as determined e.g. by molecular dynamics. The corresponding length scale parameter will in general be proportional to, but not identical with the Burgers vector length $b$. The corresponding stress field is denoted as $\Bsigma^{\textrm{g,b}}(\Br)$. As $\delta_{\textrm{p}} \gg l_{\textrm{b}}$, this field will typically exhibit much larger values near the dislocation core, while its behavior for $r \gg \delta_{\textrm{p}}$ is the same as for the field $\Bsigma^{\textrm{g},\delta_{\textrm{p}}}(\Br)$. 
\end{itemize}
The global solution is now constructed by simple superposition of these three fields in the form
\begin{equation}
	\Bsigma(\Br) = \Bsigma^{\textrm{p}}(\Br,\delta_{\textrm{p}}) + \Bsigma^{\textrm{g}}(\Br) \quad,\quad
	\Bsigma^{\textrm{g}}(\Br) = \Bsigma^{\textrm{g,b}}(\Br) - \Bsigma^{\textrm{g},\delta_{\textrm{p}}}(\Br).
	\label{eq_multiscale}
\end{equation}
Thus, the peridynamic stress field is corrected by a stress field evaluated using gradient elasticity, which is the {\em difference} between the fields $\Bsigma^{\textrm{g,b}}(\Br)$ and $\Bsigma^{\textrm{g},\delta_{\textrm{p}}}(\Br)$. The rationale behind this procedure is as follows: (i) As can be seen from Eq. (\ref{mindlinstress}), taking the difference of two gradient elasticity stress fields with different length scales eliminates the slowly decaying classical elasticity solution. The difference is composed of modified Bessel functions only which decay exponentially as functions of $r/b$ and $r/\delta_{\textrm{p}}$. As a consequence, on scales well above the horizon $\delta_{\textrm{p}}$, the field $\Bsigma^{\textrm{g}}$ is exponentially small and can be neglected. Thus, unless the dislocation is within about one horizon from the surface, the near-surface stress field is dominated by the peridynamic solution which correctly accounts for the surface boundary conditions. (ii) Within a distance of the order of $\delta_{\textrm{p}}$ from the dislocation core, the fields $\Bsigma^{\textrm{p}}$ and $\Bsigma^{\textrm{g},\delta_{\textrm{p}}}$ are practically identical (see \figref{fig:comparison}) and their difference is negligibly small. The overall stress field is thus completely controlled by $\Bsigma^{\textrm{g,b}}$, which has the correct behavior at the dislocation core. 
\begin{figure*}[h]
	\includegraphics[width=\textwidth]{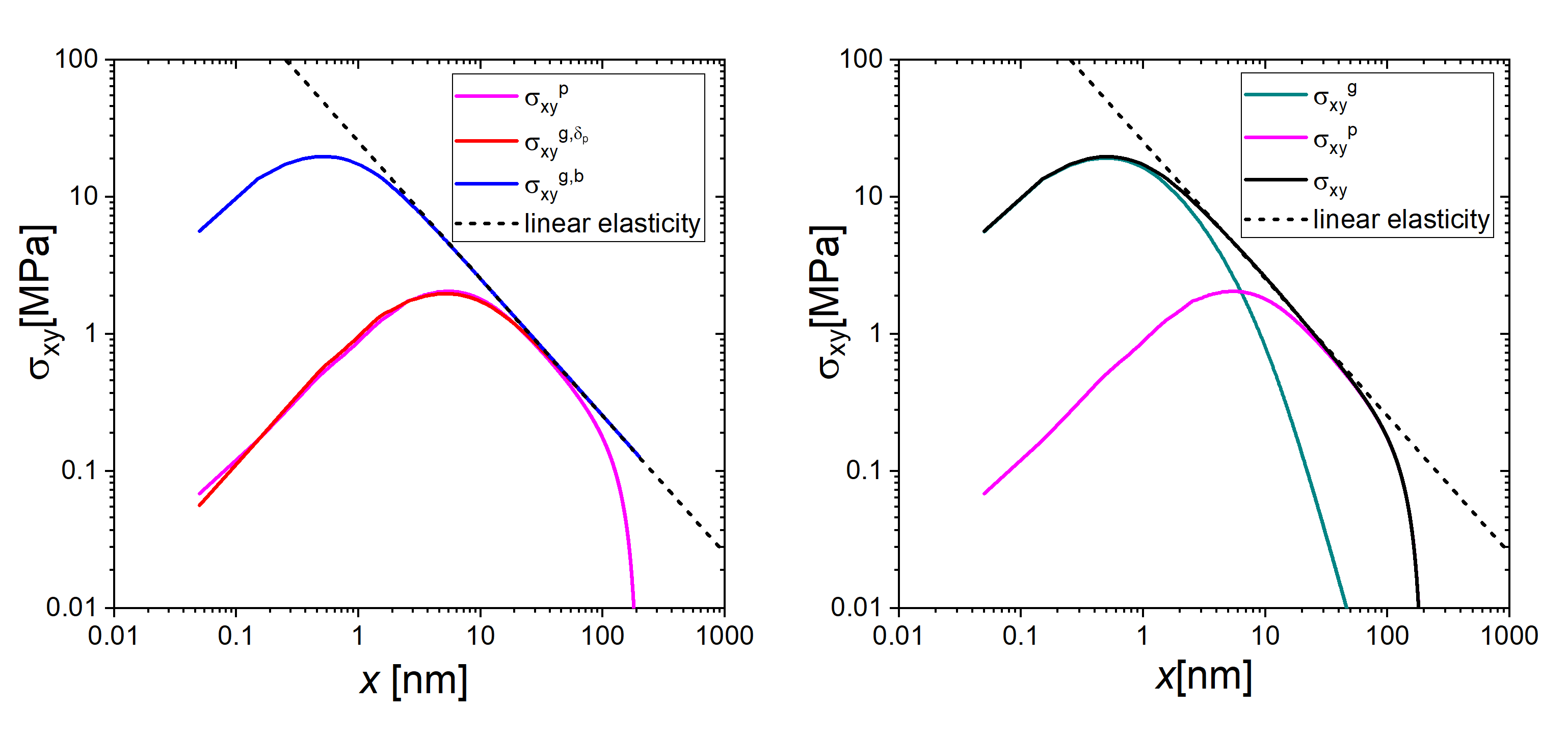}
    \caption{Construction of a hybrid dislocation model, Burgers vector $b = \SI{0.3}{\nano \metre}$, horizon of the peridynamic solution $\delta_{\textrm{p}} =  \SI{3}{\nano \metre}$, the length scale of the 'inner' gradient solution $\Bsigma^{\textrm{g,b}}$ is $l_{\textrm{b}} = \SI{0.3}{\nano \metre}$, all other parameters are as in Fig. \ref{fig:comparison}, shown is the behavior of the shear stress component $\sigma_{xy}$ on the positive $x$-axis; left: fields $\sigma^{\textrm{p}}_{xy}$,$\sigma^{\textrm{g},\delta_{\textrm{p}}}_{xy}$ and $\sigma^{\textrm{g,b}}_{xy}$, right: $\sigma^{\textrm{g}}_{xy}, \sigma^{\textrm{p}}_{xy}$ and $\sigma_{xy}$; note that the sample surface is located at $x = \SI{200}{\nano \metre}$.\label{fig:hybriddis}}
\end{figure*}
The construction is illustrated in \figref{fig:hybriddis}, showing in double-logarithmic scale the behavior of the $\sigma_{xy}$ shear stress component on the positive $x$-axis. On the left we see the three component fields with their different asymptotic behavior at large and small $x$, on the right the peridynamic stress field and the overall gradient correction together with the hybrid stress field that arises from the superposition. 
\begin{figure*}[h]
	\includegraphics[width=\textwidth]{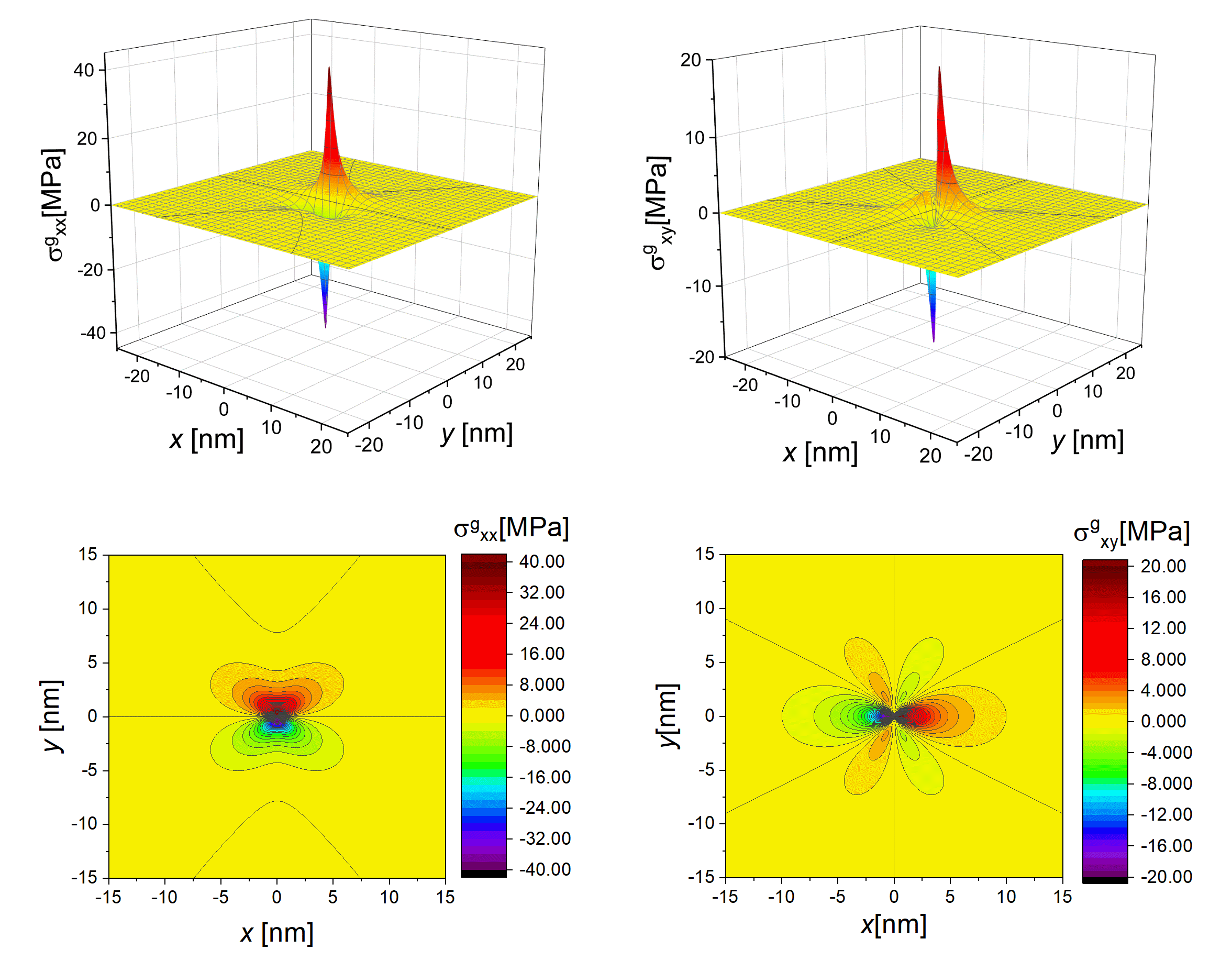}
	\caption{Angular dependence of the short-range correction function $\Bsigma^{\textrm{g}}$ as evaluated for the parameters of \figref{fig:hybriddis}; left: stress component $\sigma^{\textrm{g}}_{xx}$, right: stress component $\sigma^{\textrm{g}}_{xy}$.}
	\label{fig:gradcorr}
\end{figure*}
The angular dependence of the gradient correction field $\Bsigma^{\textrm{g}}(\Br)$ is illustrated in \figref{fig:gradcorr}. It is clearly seen that the correction field becomes negligibly small above the scale of a few horizons of the peridynamic field (here: $\delta_{\textrm{p}} = \SI{3}{\nano \metre}$). While the overall morphology matches that of the classical solution, it is noted that the angular dependency of the correction field exhibits some deviations from the classical expectation, for instance, the location of the zero-field level lines in \figref{fig:gradcorr} indicates that the stable position of close edge dislocation dipoles is under 30 degrees rather than under 45 degrees as expected according to the classical solution.  

\section{Discussion and Conclusions}

Combining peridynamics and gradient elasticity allows to construct a two-scale structure model of a dislocation which uses peridynamics for evaluating dislocation fields that correctly account for surface boundary conditions, and combines this evaluation with a short-range correction to account for the high stress fields close to the dislocation.  

The presented results provide all necessary 'ingredients' for constructing a plasticity model. Calculating the evolution of the plastic distortion is straightforward once the dislocation velocity is known: For the geometry of the edge dislocation, the plastic distortion rate fulfils the relation $\dot{\Bbetapl} = \dot{\gamma} \left[ \Be_x \otimes \Be_y \right]$, where the scalar slip rate is related to the Burgers vector distribution via $\dot{\gamma} = b v \phi(\Br)$ where $v$ is the scalar dislocation glide velocity. The temporal change of the plastic distortion gives, in turn, according to Eq. (\ref{bonddef}) the evolution of the bond structure due to the plastic deformation. The driving forces that make the dislocation move and cause multiple dislocations to interact are straightforward to evaluate since, as shown by Po et al. \cite{po2014singularity}, in Helmholtz-type gradient elasticity the Peach–Koehler force retains its classical form in the sense that it involves only the Cauchy stress acting on the dislocation line -- which we have shown how to compute. Moreover, if Helmholtz gradient elasticity is used to compute the short-distance correction fields, the approach is straightforward to generalize to three-dimensional systems of curved dislocation lines. 

There are limitations. For a dislocation within distance $\delta_{\textrm{p}}$ or less from the surface, the computation becomes unreliable because the correction stress $\Bsigma^{\textrm{g}}(\Br)$ does not account for the surface boundary conditions; as a consequence, dislocation interactions with the surface are under-estimated at short distances from the surface. This problem exists, in one form or another, in all approaches which use a continuum simulation method to provide surface corrections to bulk dislocation fields, e.g., in the classical approach of Needleman and Van der Giessen \cite{van1995discrete} where bulk stress fields are corrected by the FEM solution of a modified surface boundary value problem, the problem becomes manifest on scales below the resolution of the finite element mesh. Similarly, in the recently proposed method of Dong et. al. \cite{dong2022peridynamic}, where the finite element mesh is replaced by a peridynamic simulation approach, inaccuracies arise once the dislocation-surface spacing gets on the order of the peridynamic horizon. How to correct potential near-surface artefacts remains an important question for further work on developing the present approach.

\section*{Declarations}
\subsection*{Competing interests}
  The authors declare that they have no competing interests.
  
\subsection*{Author's contributions}
J.R. performed peridynamic simulations and data analysis, M.Z. performed theoretical calculations and drafted the manuscript. The manuscript was edited and approved jointly by all authors. 

\subsection*{Funding}
This work was funded by the Deutsche Forschungsgemeinschaft (DFG, German Research Foundation) under projects 377472739/GRK 2423/1-2019 and Za171/13-1. The authors gratefully acknowledge this support. 

\subsection*{Acknowledgements}
Not applicable

\subsection*{Availability of data and materials}
Not applicable

\bibliographystyle{elsarticle-num} 
\bibliography{periref}   

\newpage

\appendix
\setcounter{equation}{0}
\setcounter{figure}{0}
\setcounter{table}{0}
\renewcommand{\theequation}{\Alph{section}.\arabic{equation}}

\end{document}